\documentclass[journal]{IEEEtran}

\usepackage{epstopdf}
\ifCLASSINFOpdf
   \usepackage[pdftex]{graphicx}
   \graphicspath{{C:../pdf/}{C:../jpeg/}}
   \DeclareGraphicsExtensions{.pdf,.jpeg,.png}
\else
  \usepackage[dvips]{graphicx}
  \graphicspath{{C:../eps/}}
  \DeclareGraphicsExtensions{.eps}
\fi
\usepackage{amsmath}
\makeatletter

\newcommand{\Rmnum}[1]{\expandafter\@slowromancap\romannumeral #1@}
\makeatother
\usepackage{bm}
\usepackage{colortbl}
\usepackage{makecell}
\usepackage{multirow}
\usepackage{amssymb}
\usepackage{algorithm}
\usepackage{algorithmic}
\usepackage{hyperref}
\usepackage{threeparttable}
\usepackage{array}
\usepackage{float}
\usepackage{amsmath}
\usepackage{subfigure}
\usepackage{changepage}
\usepackage{footnote}

\usepackage{color}
\usepackage{cite}
\begin{document}

\title{Distributed Cooperative Driving in Multi-Intersection Road Networks}%

\author{Huaxin Pei, Yi Zhang,~\IEEEmembership{Member,~IEEE}, Qinghua Tao, Shuo Feng, Li Li,~\IEEEmembership{Fellow,~IEEE}
\thanks{Manuscript received in September 19, 2020; This work was supported in part by National 135 Key R\&D Program Projects under Grant 2018YFB1600600 and National Natural Science Foundation of China under Grant 61673233. (Corresponding author: \emph{Li Li})}
\thanks{H. Pei is with the Department of Automation, Tsinghua University, Beijing China, 100084 (e-mail: phx17@mails.tsinghua.edu.cn).}
\thanks{Y. Zhang is with the Department of Automation, BNRist, Tsinghua University, Beijing 100084, and Berkeley Shenzhen Institute (TBSI) Shenzhen 518055, China, and also with Jiangsu Province Collaborative Innovation Center of Modern Urban Traffic Technologies, Nanjing 210096, China (e-mail: zhyi@tsinghua.edu.cn).}
\thanks{Q. Tao is with the Department of Automation, Tsinghua University, Beijing China, 100084 (e-mail: taoqh14@tsinghua.org.cn).}
\thanks{S. Feng is with the Department of Civil and Environmental Engineering, University of Michigan, Ann Arbor, MI 48109 USA(e-mail:fshuo@umich.edu).}
\thanks{L. Li is with the Department of Automation, BNRist, Tsinghua University, Beijing, China, 100084 (e-mail: li-li@tsinghua.edu.cn).}}


\maketitle

\begin{abstract}
Cooperative driving at isolated intersections attracted great interest and had been well discussed in recent years. However, cooperative driving in multi-intersection road networks remains to be further investigated, because many algorithms for isolated intersection cannot be directly adopted for road networks. In this paper, we propose a distributed strategy to appropriately decompose the problem into small-scale sub-problems that address vehicle cooperation within limited temporal-spatial areas and meanwhile assure appropriate coordination between adjacent areas by specially designed information exchange. Simulation results demonstrate the efficiency-complexity balanced advantage of the proposed strategy under various traffic demand settings.
\end{abstract}

\IEEEpeerreviewmaketitle
\section{Introduction}
\IEEEPARstart{N}{ew} potential technologies in vehicle systems have shown outstanding performance in ensuring safety, improving efficiency, and saving energy. Electric vehicle technology has great potential in reducing fuel consumption and air pollution \cite{5940409,5940579}. The advanced sensing systems \cite{0Embedded} and algorithms \cite{5940405} enhance the perception ability of smart vehicles and lead to the birth of automated vehicles \cite{5940562,8884676}, which can effectively reduce collisions in transportation systems. Further, with the aid of vehicle-to-everything (V2X) communication technologies, automated vehicles evolve into connected and automated vehicles (CAVs) \cite{6823640,7934129} and they can share the driving states with adjacent vehicles, schedule the movements of neighboring vehicles to move safely and efficiently, and thus improve traffic safety and efficiency. Such techniques are usually called cooperative driving and have received increasing interest recently \cite{1,2,25,6,8,9,10,11,YANG2021102918}.
\par Most existing studies about cooperative driving focus on coordinating the vehicles within local small areas, which can be regarded as \emph{\textbf{area-wide cooperative driving}}. In these studies, we usually emphasize how to schedule the movements of vehicles that move in different directions within the conflict area so that these vehicles can pass through the conflict area efficiently without collisions \cite{3,5,7,4}. As pointed out in \cite{1,4,13}, the key problem of cooperative driving is to determine the optimal sequence of vehicles passing through the conflict area.
\par In this paper, we tackle the cooperative driving problem from the view of road network which contains multiple conflict areas, and thus it can be regarded as \emph{\textbf{network-wide cooperative driving}}. Although many useful algorithms had been proposed to promote cooperative driving at isolated conflict areas \cite{14,15,16}, there still exist challenges for cooperative driving in road networks.
\par Different from the isolated conflict areas, each vehicle needs to pass through multiple conflict areas in the road network and thus it results in extremely complicated interactions between vehicles. More seriously, as pointed out in \cite{17}, it may generate the causality cycles in the process of planning trajectory for vehicles, where the planning results of the vehicles around different conflict areas affect each other mutually so that it leads to failures when each vehicle plans its ultimately optimal trajectory \cite{17}. In such case, a few studies directly deal with all vehicles in the road network and formulate a large-scale planning problem, where each conflict between vehicles introduces a binary variable to mathematically describe vehicle sequence at conflict areas \cite{22,23}. It leads to high computational complexity and makes the centralized planning problem intractable. This causes a possible shift in solving network-wide cooperative driving problem from centralized approach to decentralized or distributed approach to guarantee computational efficiency.
\par In fact, the decentralized or distributed approaches have already been discussed in area-wide cooperative driving problem. In \cite{26} and \cite{31}, a decentralized framework was proposed to implement energy-optimal trajectory planning for the vehicles around a signal-free intersection. Xu \emph{et al.}\cite{28} provided a distributed cooperation strategy to resolve all conflicts produced at unsignalized intersections. In \cite{27}, a hierarchical-distributed coordination structure is established for intersections to improve traffic efficiency. In these strategies, each vehicle is responsible for planning its own trajectory according to the movements information of neighbouring vehicles, which can realize on-line computation at isolated conflict areas.
\par Similar to the aforementioned strategies in area-wide cooperative driving, the latest studies have tried to utilize the decentralized or distributed ideas in network-wide cooperative driving problem, e.g., \cite{20,21,29}. In these studies, the algorithms for area-wide cooperative driving are independently applied in each single conflict area of the road network, while no coordination between adjacent conflict areas. It has a poor performance in improving traffic efficiency especially when the adjacent conflict areas are close to each other, since the conflict areas affect each other mutually in a road network. Thus, distributed or decentralized cooperative driving in road networks still needs to be further studied with respects to the reduction of computational complexity and also the improvement of traffic efficiency.
\par It is worth emphasizing that the concepts of ``\emph{distributed}" and ``\emph{decentralized}"  are indistinguishable in area-wide cooperative driving problem in many studies. However, we prefer to adopt ``\emph{distributed}" to describe the nature of network-wide cooperative driving problem, because the road network needs to be divided into different areas in spatial dimension and there exists information exchange between adjacent areas in network-wide cooperative driving.
\par Therefore, in this paper, we propose a distributed strategy for network-wide cooperative driving. The key idea is to decompose the original large-scale problem into several small-scale sub-problems within limited temporal-spatial areas. Then, each vehicle is sequentially incorporated into different sub-problems along with time to realize the target of passing the whole road network, so that the vehicles around different conflict areas can be coordinated separately. Thus, the proposed distributed strategy can eliminate the causality cycles and then decrease the computational complexity.
\par In addition, we design an appropriate prediction-based coordination strategy by utilizing the information of vehicles from adjacent areas to assure cooperation between adjacent areas and thus guarantee the traffic efficiency. It leads to the current coordination results of each area better adapted to the future traffic flows. Although the distributed strategy theoretically leads to a local optimal solution, the proposed strategy shows predominant advantages over the existing cooperative driving strategy regarding both traffic efficiency and computational complexity through comparison simulations under various traffic demand settings. It demonstrates the promising performance of the distributed strategy in network-wide cooperative driving problem.
\par The proposed distributed strategy works well in road networks, and the mechanism behind its effectiveness can be traced to the following reasons. In general, the vehicle has the priority to pass through the conflict area when it is close to the area, so that the vehicles around adjacent conflict areas have a trivial influence on the trajectory planning for those vehicles close to current conflict area. Therefore, the vehicles close to the current conflict area can omit the influence of those vehicles which do not pass through the upstream conflict areas, i.e., the vehicles around different conflict areas can be tackled separately. It leads to that the causality cycles are indirectly eliminated. In our proposed strategy, besides the vehicles around the conflict area, we also incorporate the information of vehicles in the sections connecting adjacent conflict areas into each sub-problem to implement the prediction-based coordination strategy, which assures the cooperation between areas and then guarantees traffic efficiency of the whole network.
\par To better present our findings, the rest of this paper is organized as follows. \emph{Section II} presents the typical road network scenario and formulates the trajectory planning problem of cooperative driving in road networks from the perspective of operational research. \emph{Section III} proposes a distributed strategy to attack the problem of cooperative driving in road networks. Then, we provide simulation results in \emph{Section IV}. Finally, conclusion and further works are presented in \emph{Section V}.
\begin{figure}[t]
\centering
\includegraphics[width=3.3in]{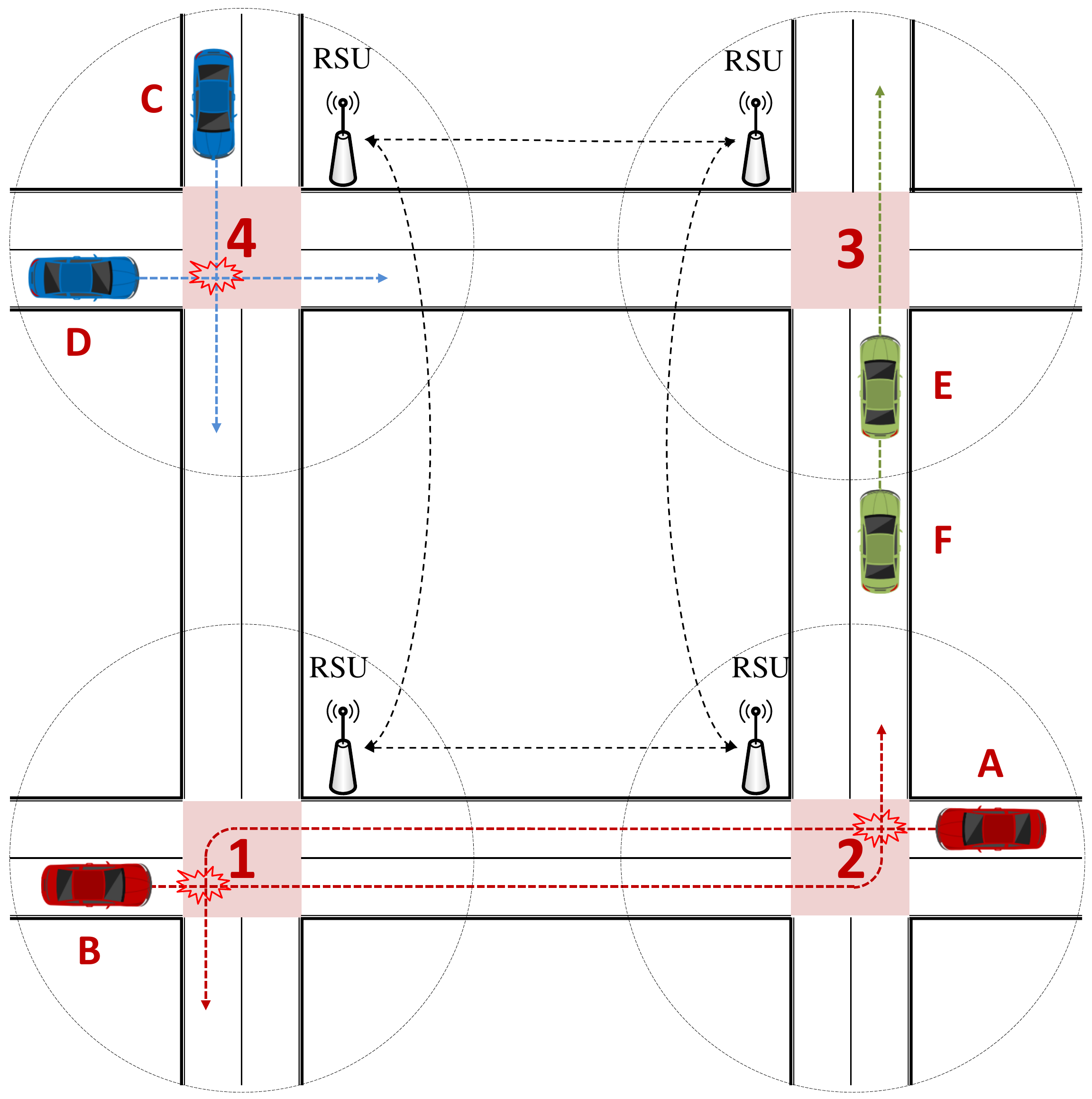}
\caption{A typical road network with four intersections.}
\label{fig1}
\end{figure}
\section{Problem Presentation}
\par In this section, we will present the typical road network scenario and formulate the trajectory planning problem of coordinating all vehicles in multi-intersection road networks.
\subsection{Scenario and Notations}
\par In this paper, we take a typical road network scenario with four signal-free intersections as an illustration to explain our method; see Fig. \ref{fig1}. Other road networks could be handled in a similar way. The red area is the conflict area of the corresponding intersection, where the vehicles from different directions may collide. Each intersection is assigned with a road-side unit (RSU) as the local controller to schedule vehicles within its control range. In addition, the adjacent intersections can exchange the traffic information with each other to realize the cooperation between different conflict areas. We denote the set of intersections index as $I$, $I=\{1,2,3,4\}$.
\begin{table*}
\renewcommand{\arraystretch}{1.2}
\centering
\caption{The Nomenclature List}
\centering
\begin{tabular}{c|l}
\Xhline{1.2pt}
{\bfseries Variables} & {\bfseries Notations}\\
\Xhline{1.2pt}
$I$ & The set of intersection index.\\

$CAV_{(i,j)}$ & The $j^{th}$ vehicle of the intersection $i$.\\

$v_c$ & The constant velocity of vehicles in the sections outside control range.\\

$J_{(i,j)}$ & The delay of $CAV_{(i,j)}$ to pass through the whole road network.\\

$J_{(i, j)}^{k}$ & The delay of $CAV_{(i,j)}$ to pass through intersection $k$.\\

$t_{\text{assign},(i,j),k}$ & The time assigned to $CAV_{(i,j)}$ to enter the conflict area of intersection $k$.\\

$t_{\text{min},(i,j),k}$ & The minimal arrival time of $CAV_{(i,j)}$ to enter the conflict area of intersection $k$.\\

$\Delta_{t,1}$ & The minimal allowable safe gap used in the physical constraints.\\

$\Delta_{t,2}$ & The maximal allowable safe gap used in the collision avoidance constraints.\\

$b_{(i,j),(i',j')}$ & The binary variable introduced in collision avoidance constraints.\\

$\emph{M}$ & The positive and sufficiently large number.\\

$v_{\text{max}}$,$a_{\text{max}}$ & The maximal velocity and acceleration of vehicles, respectively.\\

\Xhline{1.0pt}
$l_c$, $l_r$ & The length of intersection legs and road segments after dividing the road network, respectively.\\
$\omega_1$, $\omega_2$ & The weight variable used in the objective function of the sub-problems.\\
$T$ & The terminal time of the whole planning process.\\
$\Delta_T$ & The time between two consecutive optimizations of time-driven rolling horizon optimization mechanism.\\
$t_p$ & The time when the last optimization was triggered.\\

$S_{c,i}$, $S_{r,i}$ & The set of driving states of vehicles in intersection area $i$ and on the road segments, respectively.\\
\Xhline{1.2pt}
\end{tabular}
\end{table*}
\par
\begin{figure}[h]
\centering
\includegraphics[width=3.5in]{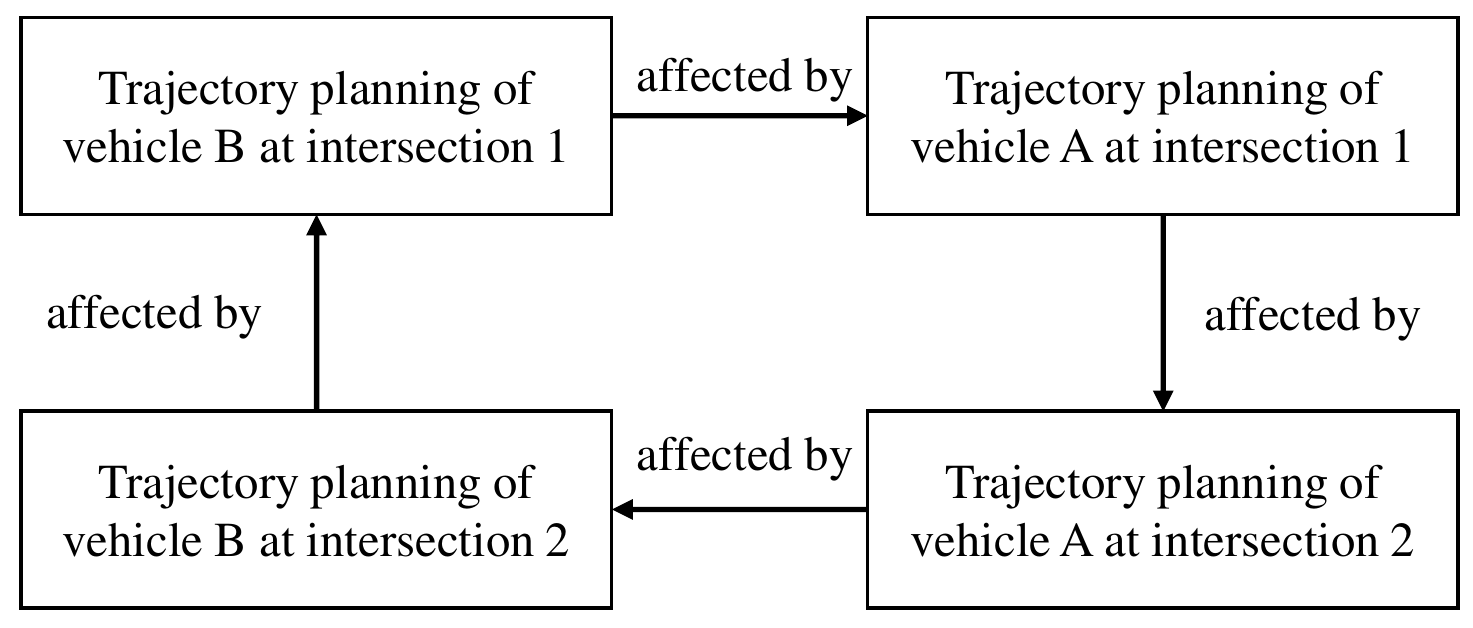}
\caption{A diagram of causality cycles generation.}
\label{fig2}
\end{figure}

 Different from isolated intersections, each vehicle needs to pass through multiple conflict areas in the road network. Thus, the conflict relations between vehicles are more complicated than that of isolated intersections. Besides the conflicts between vehicles at the same intersection, e.g., vehicle C and D, there also exist the conflicts between the vehicles at different intersections, which may generate the so called causality cycles. For instance, vehicle A will conflict with B at both conflict area 1 and conflict area 2. When planning the trajectory of vehicle B at intersection 1, it has to consider the trajectory of vehicle A at intersection 1. Then, the trajectory planning of vehicle A at intersection 1 will lead to the planning for the trajectory of vehicle A at intersection 2, which is affected by the trajectory of vehicle B at intersection 2. Thus, we can find that the trajectory planning process of vehicle A and B will generate a causality cycle, as shown in Fig. \ref{fig2}. It leads to extremely complicated interactions between vehicles and makes the planning problem intractable.
\par In addition, several reasonable assumptions about the studied scenario are added as follows: a) all vehicles are connected and automated (CAVs); b) lane-change behavior is not allowed; c) the vehicles in the sections outside the control range move with a constant velocity $v_c$, e.g., vehicle F. We adopt $CAV_{(i,j)}$ to denote the $j^{\text{th}}$ vehicle of the intersection $i$. Additionally, the main notations are shown in TABLE I.
\subsection{Trajectory Planning Problem}
\par In this sub-section, we will formulate the trajectory planning problem to schedule the vehicles in road networks from the perspective of operational research.
\vspace{0.1cm}
\par\noindent\emph{(1) Decision Variable}
\par
\begin{figure}[h]
\centering
\includegraphics[width=3.2in]{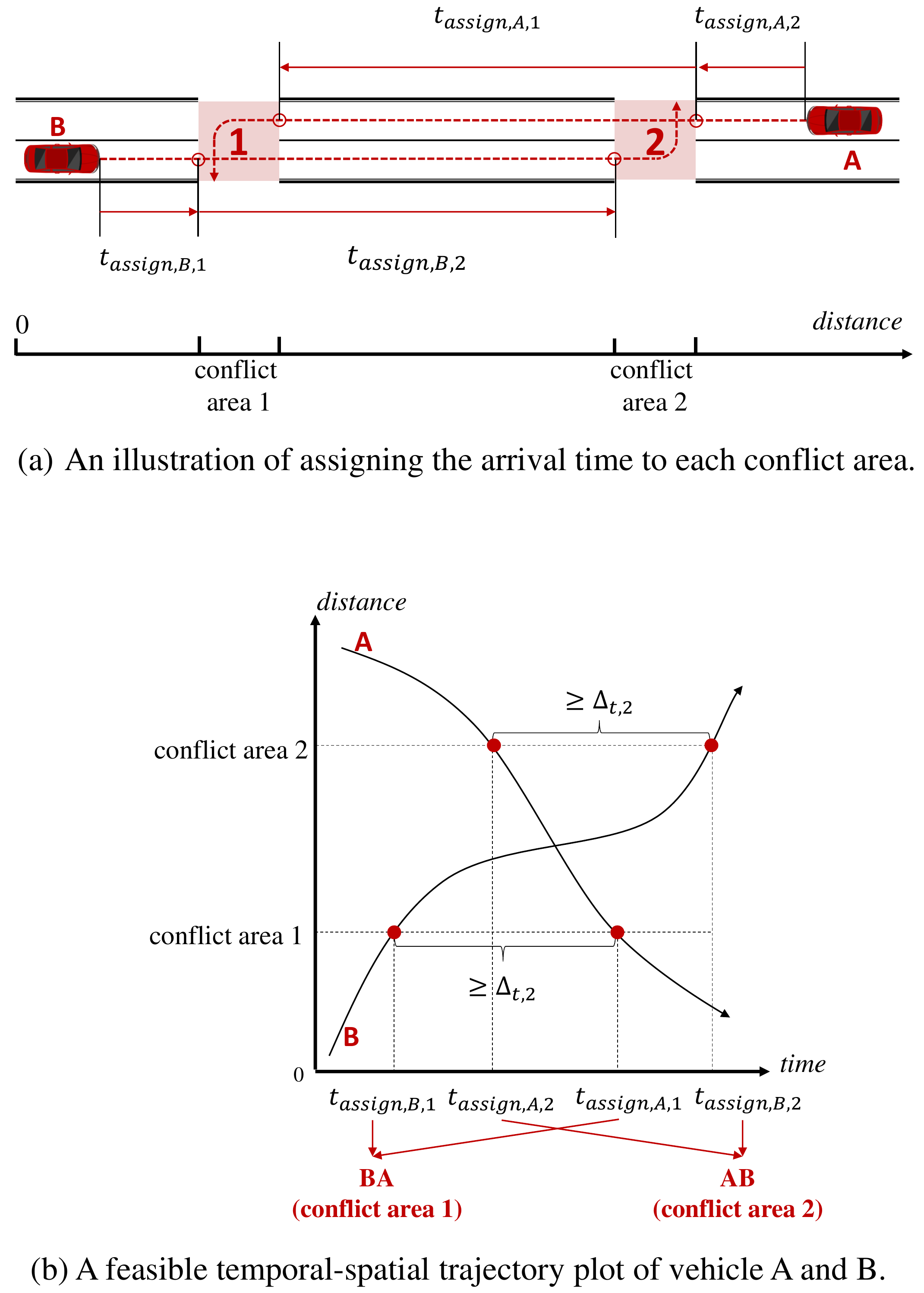}
\caption{An illustration of resolving conflicts between vehicle A and B.}
\label{fig3}
\end{figure}

 As discussed in \cite{4}, the trajectory planning in cooperative driving can be divided into two steps to decrease the problem complexity. Firstly, we need to optimize the sequence of vehicles at each conflict area, and then it is easy to derive the required trajectory in reverse according to the sequence of vehicles \cite{14,21,30}. Therefore, in this paper, we still focus on the sequence optimization at each conflict area, which can be realized through optimizing the arrival time to conflict areas for vehicles. Thus, the arrival time assigned to vehicles $t_{\text{assign},(i,j),k}$ is adopted as the decision variable of the trajectory planning problem, where $t_{\text{assign},(i,j),k}$ denotes the arrival time assigned to $CAV_{(i,j)}$ to enter conflict area $k$. As shown in Fig. \ref{fig3}(a), each vehicle will be assigned the arrival time to enter the conflict areas on its route.
\par From the trajectory plot of vehicle A and B in Fig. \ref{fig3}(b), we can find that a feasible vehicle sequence at each conflict area can be obtained via assigning the arrival time to each vehicle and then it derives the required trajectory in reverse according to the assigned time to resolve all conflicts between vehicles. Note that the causality cycle is eliminated as long as the arrival time assigned to each vehicle is feasible. It needs to impose many constraints of the trajectory planning problem, which will be introduced later.
\vspace{0.1cm}
\par\noindent\emph{(2) Objective Function}
\par According to the assigned arrival time of each vehicle, it is easy to obtain the delay of vehicles at each intersection, i.e.,
\begin{equation}
\label{1}
    J_{(i, j)}^{k}=t_{\text {assign},(i, j), k}-t_{\min ,(i, j), k},
\end{equation}
\par\noindent where $J_{(i,j)}^{k}$ denotes the delay of $CAV_{(i,j)}$ when passing through intersection $k$, and $t_{\min ,(i, j), k}$ denotes the minimal arrival time of $CAV_{(i,j)}$ to enter the conflict area of intersection $k$, which satisfies the vehicle dynamics constraints and can be obtained \cite{15}.
\par To improve traffic efficiency, cooperative driving in road networks aims to minimize the average delay of vehicles passing through the road network. Since many studies have discussed how to select appropriate driving routes for vehicles in a road network \cite{20,AN2020102777}, the problem of route planning is not addressed in this paper, i.e., we assume that the driving routes of vehicles are pre-determined. Thus, the delay of vehicle passing through the whole road network is the accumulated delay produced at different sections on its route, and we can get
\begin{equation}
\label{2}
    J_{(i, j)}=\sum_{k=1}^{K} J_{(i, j)}^{k},
\end{equation}
\par\noindent where $J_{(i, j)}$ denotes the delay of $CAV_{(i,j)}$ to pass through the road network, and $K$ denotes the number of intersections on the route of $CAV_{(i,j)}$.
\par Therefore, the objective function of cooperative driving in road networks can be formulated as
\begin{equation}
\label{3}
\min _{t_{\text {assign},(i, j), k}} \frac{\sum_{i=1}^{P} \sum_{j=1}^{Q} J_{(i, j)}}{N},
\end{equation}
\par\noindent where $P$ and $Q$ denote the number of intersections in the road network and the number of vehicles around the corresponding intersection, respectively. $N$ denotes the total number of vehicles in the road network.
\begin{figure*}
\centering
\includegraphics[width=7.2in]{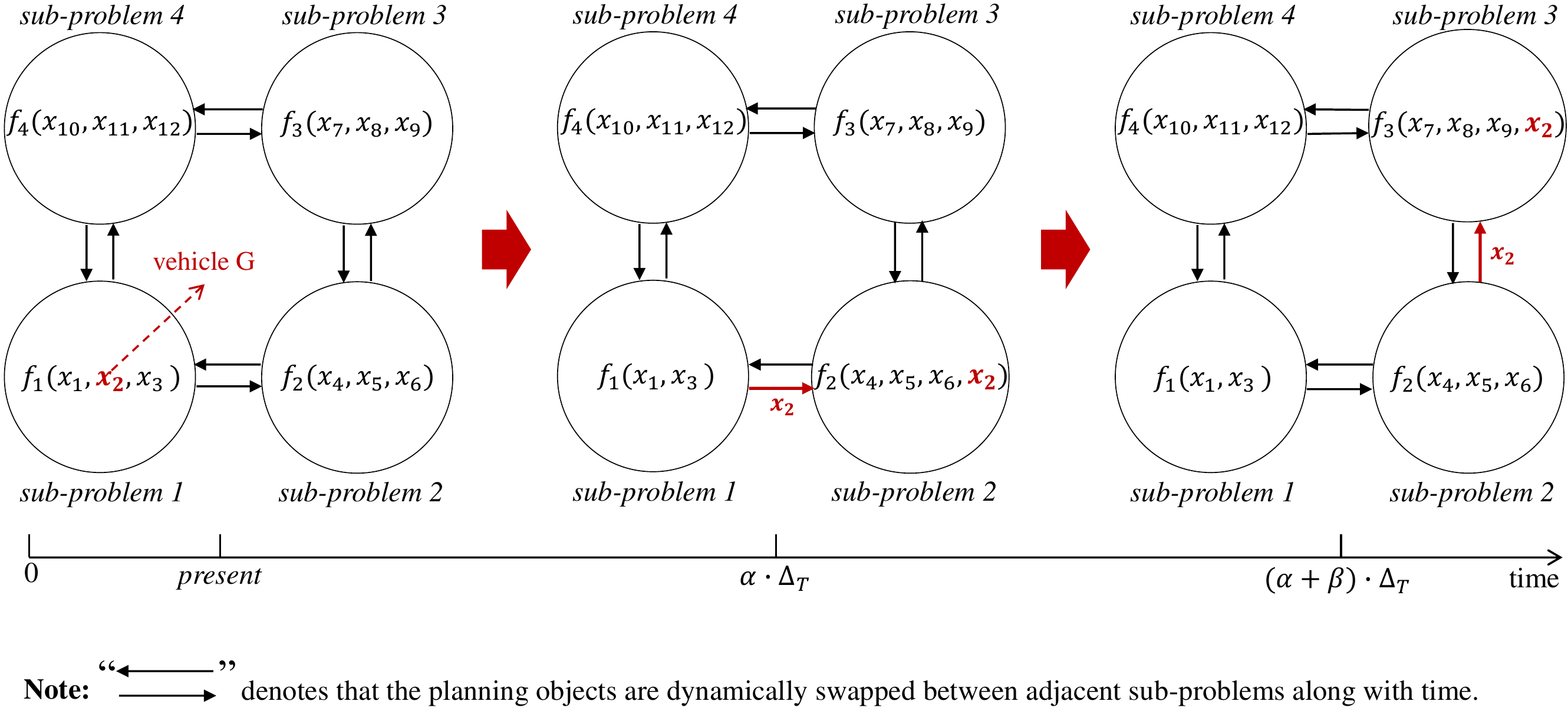}
\caption{An illustration of problem decomposition using the proposed sequential decomposition strategy ($\alpha,\beta>0$).}
\label{fig4}
\end{figure*}
\vspace{0.1cm}
\par\noindent\emph{(3) Constraints}
\par There are mainly three kinds of constraints of the trajectory planning problem including vehicle dynamics constraints, physical constraints and collision avoidance constraints. For each vehicle, these three kinds of constraints exist at all conflict areas on its route of passing through the road network and need to be resolved at once.
\par\noindent \emph{1) Vehicle Dynamics Constraints}
\par To avoid unrealistic arrival time assigned to vehicles, the assigned arrival time to enter each conflict area satisfies vehicle dynamics, i.e.,
\begin{equation}
\label{4}
    t_{\text{assign},(i, j), k} \geq t_{\min ,(i, j), k}.
\end{equation}
\par\noindent \emph{2) Physical Constraints}
\par To avoid the rear-end collision, the consecutive vehicles moving on the same lane hold the physical constraints, i.e.,
\begin{equation}
\label{5}
    t_{\text{assign},(i, j), k}-t_{\text{assign},\left(i^{\prime}, j^{\prime}\right), k} \geq \Delta_{t, 1},
\end{equation}
\par\noindent where $\Delta_{t,1}$ is the minimal safe gap for avoiding rear-end collision, and $CAV_{(i,j)}$ is physically ahead of $CAV_{(i',j')}$, such as vehicle E and F in Fig. \ref{fig1}.
\par\noindent \emph{3) Collision Avoidance Constraints}
\par To resolve the conflicts between vehicles from different directions, the vehicles should pass through the conflict area sequentially. Thus, the collision avoidance constraints are imposed by introducing the binary variables, i.e.,
\begin{equation}
\label{6}
t_{\text{assign},(i, j), k}-t_{\text{assign},\left(i^{\prime}, j^{\prime}\right), k}+M \cdot b_{(i, j),\left(i^{\prime}, j^{\prime}\right)} \geq \Delta_{t, 2},
\end{equation}
\begin{equation}
\label{7}
t_{\text{assign},\left(i^{\prime}, j^{\prime}\right), k}-t_{\text{assign},(i, j), k}+M \cdot\left(1-b_{(i, j),\left(i^{\prime}, j^{\prime}\right)}\right) \geq \Delta_{t, 2},
\end{equation}
\par\noindent where $CAV_{(i,j)}$ and $CAV_{(i',j')}$ are two vehicles from different directions and will meet at intersection $k$, such as vehicle C and D in Fig. \ref{fig1}, or vehicle A and B in Fig. \ref{fig3}(a). $\Delta_{t,2}$ is the minimal safe gap for avoiding collision between vehicles from different directions, as shown in Fig. \ref{fig3}(b). In addition, $M$ is a positive and sufficiently large number and $b_{(i,j),(i',j' )} \in \{0,1\}$. Obviously, $b_{(i,j),(i',j')}=1$ means that $CAV_{(i,j)}$ can enter the conflict area earlier than $CAV_{(i',j')}$.
\par Based on the above descriptions, the problem of sequence optimization in road networks can be mathematically formulated as a mixed-integer programming (MIP) problem, i.e.,
\begin{equation}
\label{8}
\min _{t_{\text{assign},(i, j), k}, b_{(i, j),\left(i^{\prime}, j^{\prime}\right)}} \frac{\sum_{i=1}^{P} \sum_{j=1}^{Q} J_{(i, j)}}{N},
\end{equation}
\centerline{\text {s.t.} (\ref{4})(\ref{5})(\ref{6})(\ref{7}).}
\par Note that each conflict at conflict areas yields a binary variable, which contains both the conflicts between vehicles at the same intersection and the conflicts between vehicles at different intersections illustrated in \emph{Section II-A}. Compared with cooperative driving at isolated intersections, there exist more conflicts in road networks, which makes the size of solution space soar drastically, especially when there is a large number of vehicles and intersections in the road networks. Thus, attaining a good solution in such huge searching space is an exceedingly time-consuming process by solving problem (\ref{8}) or directly adopting the existing algorithms proposed for isolated intersections, which makes problem (\ref{8}) intractable. To this end, it is necessary to propose a novel strategy to deal with the large-scale planning problem.
\section{Distributed Cooperative Driving Design}
\par In this section, we introduce a distributed cooperative driving strategy for road networks, which can decrease computational complexity and also guarantee traffic efficiency.
\par In problem (\ref{8}), each vehicle needs to plan the trajectory of passing through the whole road network in one-time planning procedure, which leads to strong coupling between adjacent areas. In fact, the vehicles around other conflict areas have a trivial influence on the trajectory planning for those vehicles close to current conflict area. Thus, the idea of ``\emph{sequential decomposition}" can be proposed to tackle the problem concerning computational burden. Specifically, we divide the road network into different areas in the spatial dimension, and then each vehicle will sequentially move into different areas on its route and plan the trajectory of the corresponding areas along with time, so that the vehicles within different areas can be coordinated separately. Therefore, the constraints of problem (\ref{8}) between adjacent conflict areas are relaxed and the causality cycles are eliminated through sequential decomposition.
\par Inspired by the above ideas, we propose a distributed strategy to decompose problem (\ref{8}) into several sub-problems within limited temporal-spatial areas. Specifically, each intersection leads to a sub-problem to schedule the vehicles within its control range. At the same time, a prediction-based coordination strategy is integrated into each sub-problem to guarantee the traffic efficiency, where the driving states of vehicles outside the control range are incorporated into the sub-problems to plan trajectories together with the vehicles within the control range, so that the current coordination results of each intersection can better adapt to the future traffic flow. Meanwhile, the optimal trajectories are reserved for the vehicles outside the control range, and then those vehicles can immediately drive with the optimal trajectories once arriving in the control range of the corresponding intersection without waiting for the next planning procedure, which can further utilize the temporal-spatial resources of traffic systems and then guarantee the traffic efficiency. As for the distributed strategy, we design that the trajectory planning procedure is triggered at a regular time interval $\Delta_T$ in the continuous traffic process, which can be regarded as \emph{time-driven rolling horizon optimization mechanism}\cite{14,16}.
\par The rest of \emph{Section-III} is organized as follows. In \emph{Section-III-A}, we propose a sequential decomposition strategy to implement problem decomposition. \emph{Section-III-B} mathematically formulates each sub-problem, and \emph{Section-III-C} introduces the solving algorithm to search a good solution of each sub-problem. Finally, we give a brief discussion on the scalability of the proposed distributed strategy in \emph{Section-\Rmnum{3}-D}.
\subsection{Problem Decomposition}
\par In this sub-section, a sequential decomposition strategy is proposed to decompose problem (\ref{8}) into several small-scale sub-problems, and then we define the coverage areas of each sub-problem through road network division.
\vspace{0.1cm}
\par\noindent \emph{(1) Sequential Decomposition Strategy}
\par To relax the constraints between adjacent conflict areas of problem (\ref{8}), we propose a sequential decomposition strategy to decompose problem (\ref{8}) into several small-scale sub-problems. Through the cooperation among these sub-problems, it is computationally efficient to realize cooperative driving in road networks. As shown in Fig. \ref{fig4}, we give an illustration to present the sequential decomposition strategy from the perspective of operational research.
\par We use $f_i(\cdot)$ to denote the decomposed sub-problem $i$. $x_i$ denotes the planning object $i$ of sub-problems, and the planning objects refer to the vehicles in the road network. The sequential decomposition strategy is summarized as following two aspects.
\par In the spatial dimension, instead of dealing with all planning objects in problem (\ref{8}), we divide the road network into different areas and each area leads to a sub-problem, so that the planning objects are assigned to different sub-problems. Thus, each sub-problem is a small-scale planning problem and these sub-problems can work synchronously to respectively deal with a limited number of planning objects.
\par In the temporal dimension, the planning objects are swapped among these sub-problems sequentially along with time, because each vehicle needs to pass through multiple conflict areas in the road network. Thus, for each planning object, the optimization target is sequentially realized in different small-scale sub-problems rather than in one-time planning procedure as in problem (\ref{8}).
\par For instance, as shown in Fig. \ref{fig1}, vehicle B needs to pass through three conflict areas in the road network. Thus, vehicle B is sequentially incorporated into three sub-problems in the distributed strategy, i.e., sub-problem 1, 2 and 3. As shown in Fig. \ref{fig4}, we use the planning object $x_2$ to denote vehicle B. Firstly, $x_2$ is optimized by sub-problem 1, and then sub-problem 2 is responsible for its optimization target when vehicle B moves into the coverage area of sub-problem 2. Finally, the planning object $x_2$ is incorporated into sub-problem 3 with time going on. Thus, the task of vehicle B passing through the whole road network is completed sequentially by the cooperation among these sub-problems.
\begin{figure}[h]
\centering
\includegraphics[width=3.5in]{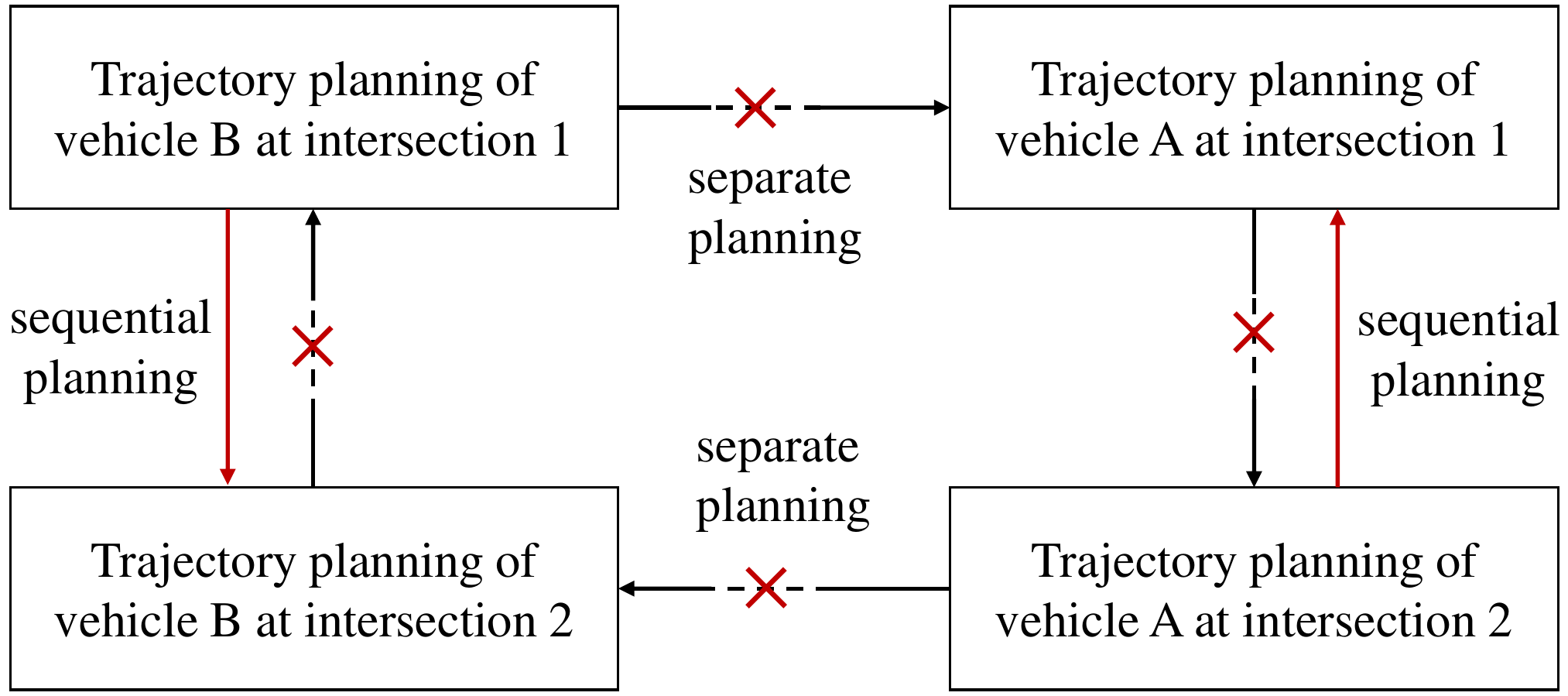}
\caption{A diagram of causality cycles elimination.}
\label{fig5}
\end{figure}
\par Based on the sequential decomposition strategy, the causality cycles are eliminated, because the vehicles around different conflict areas are separately coordinated in different sub-problems and the trajectory of each vehicle passing through the whole road network is sequentially planned in different sub-problems. As discussed in \emph{Section II-A}, vehicle A and B in Fig.\ref{fig1} may generate the so called causality cycle when implementing trajectory planning. In the proposed distributed strategy, when planning the trajectory of vehicle B at intersection 1, the trajectory of vehicle A at intersection 1 does not need to be considered. Similarly, the trajectory planning of vehicle B at intersection 2 does not affect the planning for the trajectory of vehicle A at intersection 2. In addition, the trajectory planning of vehicle A at intersection 1 and intersection 2 will be sequentially realized in different sub-problems with time going on, so does vehicle B. Thus, the causality cycle between vehicle A and B is completely eliminated, as shown in Fig. \ref{fig5}.
\vspace{0.1cm}
\par\noindent \emph{(2) Coverage Areas of Each Sub-Problem}
\par In this sub-section, we propose an appropriate rule to divide the road network into different areas to define the coverage areas of each sub-problem.
\par
\begin{figure}[t]
\centering
\includegraphics[width=3.5in]{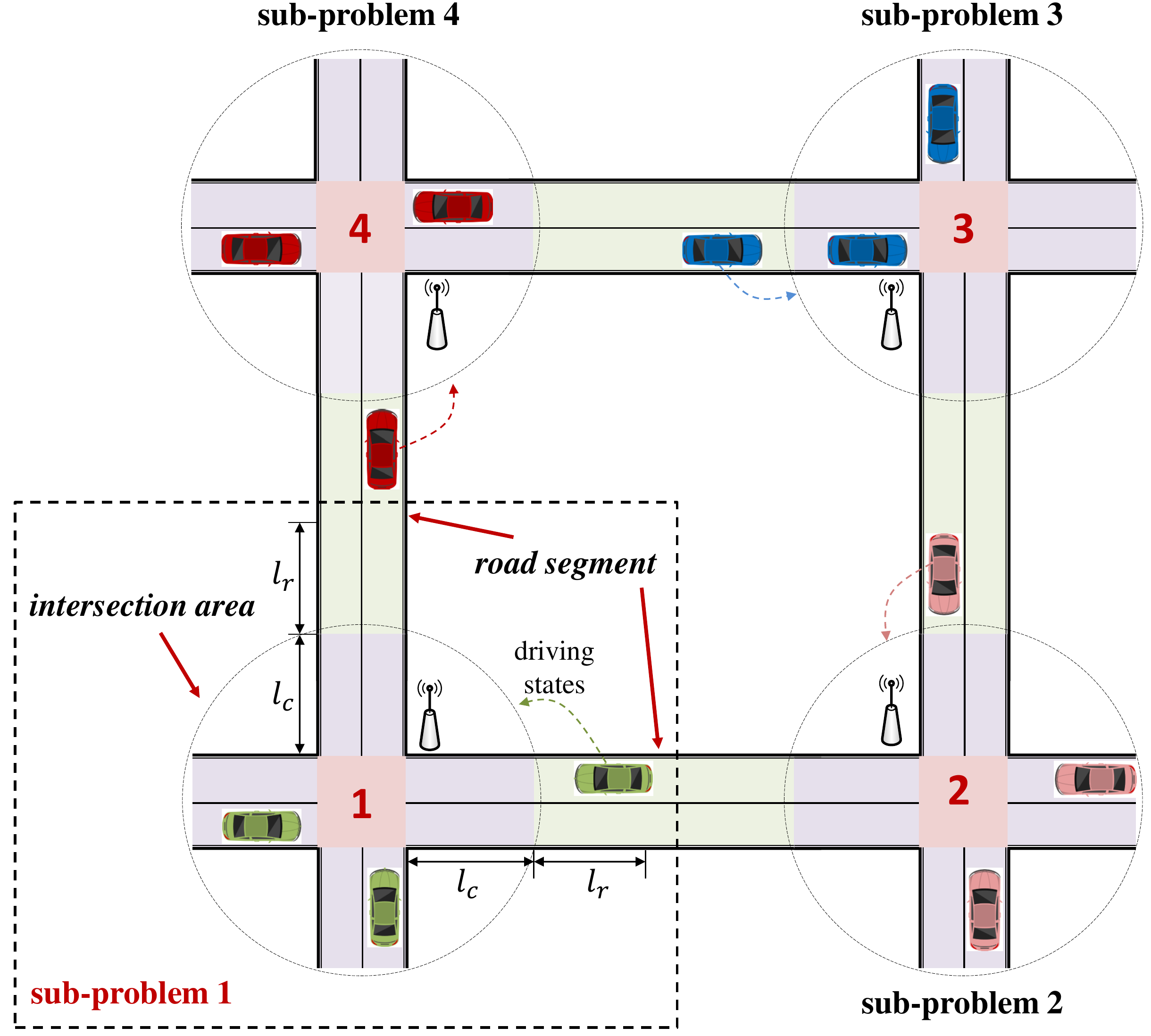}
\caption{An illustration of defining the coverage areas of each sub-problem through road network division.}
\label{fig6}
\end{figure}

 The key idea of dividing road networks is that the vehicles around different conflict areas are coordinated in different sub-problems to eliminate the causality cycles and decentralize the computational burden. Thus, the first step is to extract all conflict areas in the road network. Then, we define two kinds of basic areas covered by each sub-problem, consisting of an intersection area and the road segments. As shown in Fig. \ref{fig6}, the intersection area includes a single conflict area and the areas close to the corresponding conflict area, and the road segment is the area away from the conflict area, but connects the adjacent intersection areas.
\par The range of the intersection area $l_c$ is determined by the limited reliable communication range of vehicle-to-infrastructure communication technologies (V2I), which is similar to that of an isolated intersection in previous studies \cite{14,16}. Thus, the vehicle cooperation in each intersection area can be well resolved with the existing computationally efficient strategies proposed for isolated intersection.
\par The range of road segments that needs to be considered into each sub-problem is related to the computational resources, because the computational complexity increases with the incorporation of vehicles on road segments. In the case of sufficient computational resources, the range of the road segments should be considered as large as possible, so that each sub-problem can obtain as much information of vehicles on road segments as possible to implement prediction-based strategy to maximize traffic efficiency. In addition, in the prediction-based strategy, the intersection area must reserve the trajectory for those vehicles on road segments before they arrive in intersection area. Therefore, there is a minimal length of road segments considered into the sub-problems, i.e.,
\begin{equation}
\label{9}
    l_{r} \geq v_{c} \cdot \Delta_{T},
\end{equation}
\par\noindent where $l_r$ denotes the length of road segments that is considered into the sub-problem, as shown in Fig. \ref{fig6}. It ensures that the vehicles on the road segments have been incorporated into the corresponding sub-problem at least once in the time-driven rolling horizon optimization mechanism before they arrive in the intersection area.
\subsection{Sub-Problem Formulation}
\par In this sub-section, following the formulation procedures of problem (\ref{8}) described in \emph{Section II-B}, we can mathematically formulate each sub-problem.
\vspace{0.1cm}
\par\noindent \emph{(1) Decision Variable}
\par Similar to problem (\ref{8}), the decomposed sub-problems also aim to optimize the vehicle sequence of passing through the conflict areas. Thus, the decision variable of each sub-problem $t_{\text{assign},(i,j),k}$ is also the arrival time assigned to the vehicles within its responsibility to enter the conflict area.
\vspace{0.1cm}
\par\noindent \emph{(2) Objective Function}
\par Different from problem (\ref{8}), each sub-problem only needs to optimize the vehicle sequence at one conflict area and then obtain a limited temporal-spatial trajectory for each vehicle to pass through the corresponding intersection area. Thus, the objective of each sub-problem is to minimize the average delay of vehicles of passing through current intersection area. In addition, the driving states of the vehicles on the road segments are incorporated into each sub-problem to implement prediction-based coordination to guarantee traffic efficiency. Therefore, the objective function of sub-problem $k$ can be formulated as a weighted sum expression, i.e.,
\begin{equation}
\label{10}
\min _{t_{\text{assign},(i, j), k}} \frac{\sum_{j=1}^{Q_{1}} \omega_{1} \cdot J_{(i, j)}^{k}+\sum_{j=Q_{1}+1}^{Q_{2}} \omega_{2} \cdot J_{(i, j)}^{k}}{Q_{1}+Q_{2}},
\end{equation}
\par\noindent where $Q_1$ and $Q_2$ denote the number of vehicles in the intersection area $k$ and on the corresponding road segments, respectively. Generally, the vehicles close to the conflict areas are more likely to rank higher in the sequence than those away from the conflict area. Thus, the delays of vehicles in the intersection area and on the road segments are assigned with different weights, i.e., $\omega_1$ and $\omega_2$, which instructs the process of searching an appropriate vehicle sequence.
\par According to (\ref{10}), it can be found that each sub-problem only needs to deal with limited number of vehicles and plan the trajectory of passing one intersection for each vehicle. Thus, compared with problem (\ref{8}), there exists less number of decision variables in each sub-problem.
\vspace{0.1cm}
\par\noindent \emph{(3) Constraints}
\par The constraints of each sub-problem still include three types, i.e., vehicle dynamics constraints, physical constraints and collision avoidance constraints, as shown in (\ref{4})-(\ref{7}). For each vehicle, these three kinds of constraints only exist at current conflict area, while the conflicts produced at other conflict areas are not considered in current sub-problem. Thus, compared to problem (\ref{8}), the sub-problem is a small-scale MIP problem, i.e.,
\begin{equation}
\label{11}
\min _{t_{\text{assign},(i,j),k}, b_{(i, j),\left(i^{\prime}, j^{\prime}\right)}} \frac{\sum_{j=1}^{Q_{1}} \omega_{1} \cdot J_{(i, j)}^{k}+\sum_{j=Q_{1}+1}^{Q_{2}} \omega_{2} \cdot J_{(i, j)}^{k}}{Q_{1}+Q_{2}},
\end{equation}
\centerline{\text {s.t.} (\ref{4})(\ref{5})(\ref{6})(\ref{7}).}
\par In the above three kinds of constraints, the collision avoidance constraint is the main factor affecting the computational complexity, since the binary variables are introduced in the collision avoidance constraints. In problem (\ref{8}), not only the vehicles around the same conflict area will produce binary variables, but also the vehicles around different conflict areas. However, in problem (\ref{11}), the collision avoidance constraints of vehicles around different conflict areas are eliminated and thus the number of the binary variables in problem (\ref{11}) is extremely less than that in problem (\ref{8}). A more detailed explanation of the reduction in computational complexity is as follows.
\par We use $n_1$ to denote the total number of the conflicts between vehicles around the same intersections in the road network, e.g., vehicle C and D in Fig. 1. $n_2$ denotes the total number of the conflicts between vehicles around different intersections, e.g., vehicle A and B in Fig. \ref{fig1}. For instance, as shown in Fig. \ref{fig1}, $n_1=1$ and $n_2=2$.
\par In the road network scenario, each vehicle needs to pass through multiple conflict areas and then produces the conflicts with the vehicles around those conflict areas. Thus, $n_2 > n_1$ is usually holds especially when there are many conflict areas in the road network.
\par Each conflict introduces a binary variable to formulate the collision avoidance constraint, and all of the above collision avoidance constraints are resolved at once in problem (\ref{8}). Thus, the size of solution space of problem (\ref{8}) can be abstracted as
\begin{equation}
\label{13}
    s=2^{n_{1}+n_{2}},
\end{equation}
\par\noindent where $s$ denotes the size of solution space of problem (\ref{8}).
\par As for sub-problem (\ref{11}), the conflicts between vehicles around different conflict areas are not taken into account. Thus, the size of solution space of each sub-problem can be approximately abstracted as
\begin{equation}
\label{14}
    s^{\prime}=2^{n_{1} / P},
\end{equation}
\par\noindent where $s'$ denotes the size of solution space of each sub-problem, and $P$ denotes the number of conflict areas in the road network.
\par It can be noticed that $s'$ is significantly smaller than $s$ especially when there are many conflict areas and vehicles in the studied road network. Therefore, the computational complexity of sub-problem (\ref{11}) is extremely decreased compared with that of problem (\ref{8}).
\subsection{Sub-Problem Solving Algorithm}
\par In this sub-section, we introduce a modified Monte Carlo tree search (MCTS) to search a good solution of each sub-problem efficiently, and then we give a complete workflow of each sub-problem in the distributed strategy.
\par Directly solving the sub-problems is still a time-consuming process, even if the size of solution space is significantly smaller than problem (\ref{8}). As pointed out in \cite{16,24}, the MCTS method has shown potentials to approach an approximately optimal solution at isolated intersections within limited budgets. In this paper, we modify and extend this promising tree search method to solve problem (\ref{11}), where the vehicles around the intersection area and on the upstream road segments are assigned with different priorities rather than being treated equally. Specifically, in the search process of MCTS, each node is assigned with a score to evaluate its potential in approaching the optimal solution. The score of each node is defined as the average delay for all vehicles contained in the corresponding node \cite{16}. To instruct the search process, the derivation of score is modified with the objective function of problem (\ref{11}) to improve the search performance within a limited computational budget, i.e.,
\begin{equation}
S=\frac{\sum_{j=1}^{Q_{1}} \omega_{1} \cdot J_{(i, j)}^{k}+\sum_{j=Q_{1}^{\prime}+1}^{Q_{2}^{\prime}} \omega_{2} J_{(i, j)}^{k}}{Q_{1}^{\prime}+Q_{2}^{\prime}},
\end{equation}
\par\noindent where $Q_1^{\prime}$ and $Q_2^{\prime}$ respectively denote the number of vehicles in the intersection area $k$ and on the road segments in the corresponding node. More details about MCTS method in searching a good vehicle sequence can refer to \cite{16,24}.
\begin{figure}[t]
\centering
\includegraphics[width=3.5in]{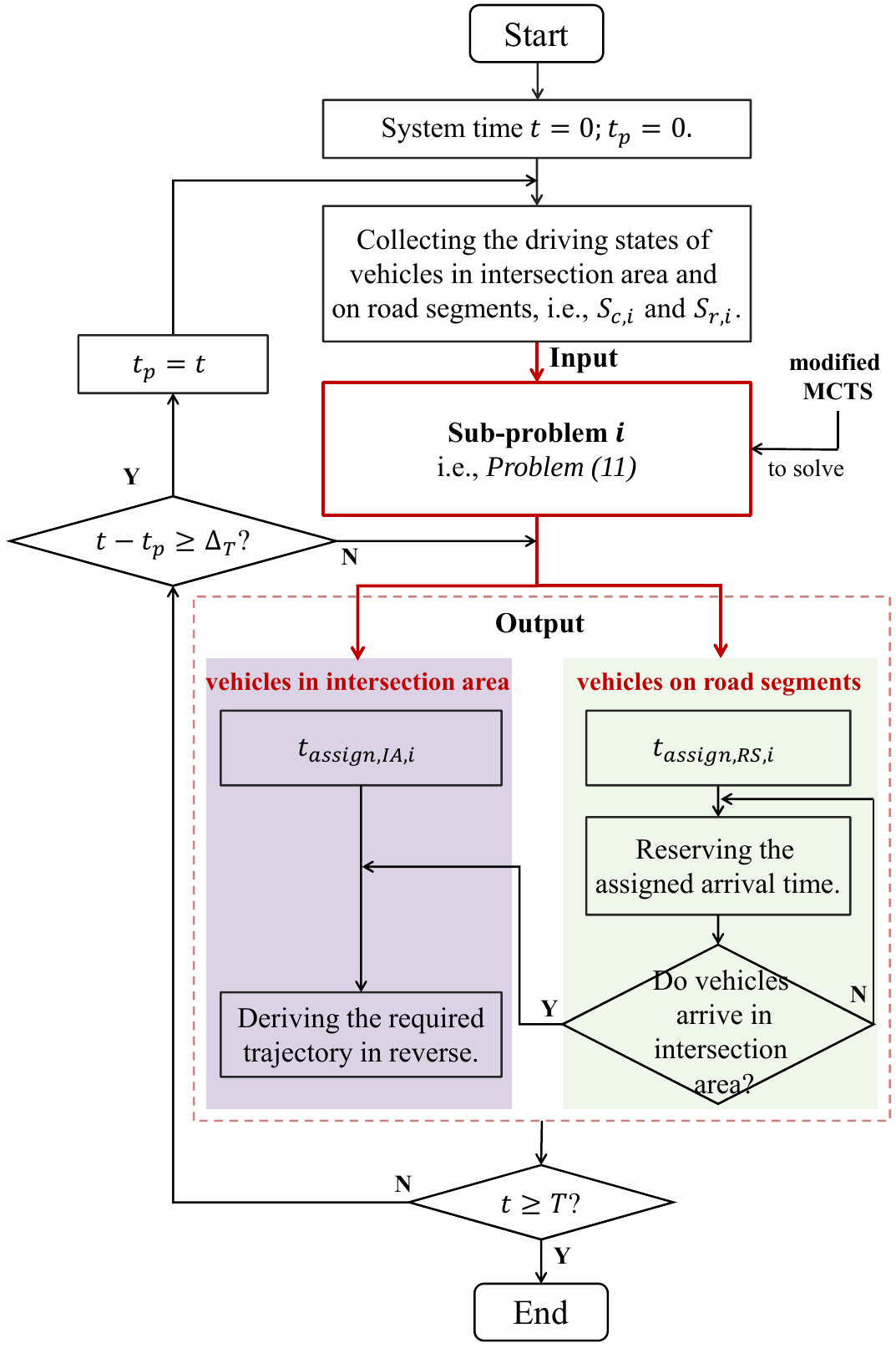}
\caption{An illustration of the workflow of sub-problem $i$.}
\label{fig7}
\end{figure}
\par After solving all sub-problems synchronously, we can obtain the arrival time assigned for the vehicles both in each intersection area and also the vehicles on the road segments to enter the corresponding conflict area, and then it is easy to derive the required trajectory in reverse according to the assigned time \cite{14,20}. As for vehicles in each intersection area, they will follow the planned trajectories, which can better adapt to the future traffic flow with employing of the driving states of vehicles on the road segments to safely and efficiently pass through the corresponding conflict area. As for vehicles on the upstream road segments, the intersection area reserves the assigned arrival time for those vehicles instead of planning the trajectory to save computational resources. Then, the trajectory of those vehicles will be re-planed according to the reserved assigned time once arriving in the intersection area without waiting for the next planning results, which can further improve traffic efficiency.
\par The workflow of sub-problem $i$ is summarized in Fig. \ref{fig7}, where $t$ is the system time and $T$ denotes the time when the whole planning process will be terminated. The planning procedure is triggered at regular time $\Delta_T$ in the time-driven rolling horizon optimization mechanism, and $t_p$ denotes the time when the last optimization was triggered. $S_{c,i}$ denotes the set of driving states (location, velocity and acceleration) of vehicles in intersection area $i$, and $S_{r,i}$ is the set of driving states of vehicles on the corresponding road segments. In addition, $t_{\text{assign},IA,i}$ and $t_{\text{assign},RS,i}$ denote the time assigned to the vehicles in intersection area and on road segments after solving sub-problem $i$, respectively.
\subsection{Scalability Discussion}
\par The proposed strategy can work well in a multi-intersection road network, because of the following reasons. Generally, the intersections in the road network are not very close to each other in space so that the vehicles around other conflict areas have a trivial influence on planning trajectory for the vehicles close to current conflict area. Intuitively, traffic management and control in a road-network scenario is a distributed problem in nature. Therefore, the coordination of vehicles around the current conflict area does not need to consider the vehicles that are far away from the current conflict area or even do not passed through the upstream conflict areas. It is entirely in time to consider the upstream vehicles when they pass through the upstream conflict area and move to the current conflict areas, and that is what the prediction-based coordination strategy does in the distributed strategy.
\par
\begin{figure}[t]
\centering
\includegraphics[width=3in]{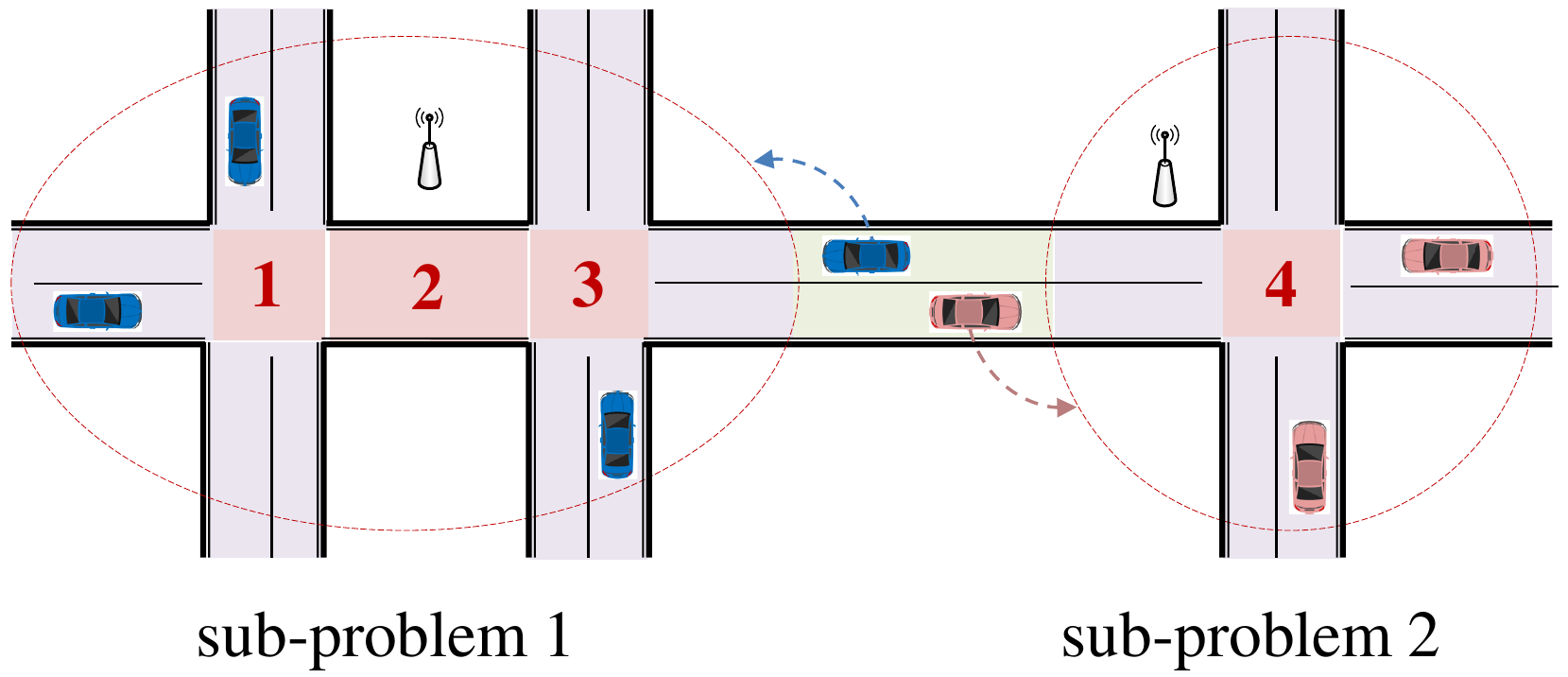}
\caption{Combining two close conflict areas as a large conflict area.}
\label{fig8}
\end{figure}

 As for the scenarios where the adjacent conflict areas are within the reliable communication range of one control unit \cite{23}, the proposed strategy can also deal with the trajectory planning problem. In such case, as shown in Fig. \ref{fig8}, the two close conflict areas need to be considered in combination, and the road segments between two adjacent conflict areas can also be regarded as a part of the combined conflict area so that the methods promoted for single intersection can be applied in such new topology, as there is no difference in nature of the methods to resolve the conflicts even if there exists a slightly more complex conflict relation of vehicles in the combined conflict area than that in an isolated conflict area.
\section{Simulation Results}
\par Three simulations are designed in this section. The first simulation evaluates the optimality and performance of the proposed strategy under a simple but common road network. The second simulation verifies the scalability of the proposed strategy in a large-scale and randomized road network under various settings. The third simulation addresses sensitivity analysis with respect to the key factors including road network geometry and parameter settings.
\subsection{Simulation Settings}
\par In following simulations, similar to \cite{14,32}, we adopt the method introduced in \cite{29} to obtain the required trajectory in reverse according to the assigned arrival time of each vehicle in the continuous traffic process. We assume that the vehicles arrive in a Poisson Process at each input lane and the average arrival rate is varied to test the performance of the proposed strategy under different traffic demands. According to \cite{15,32}, the minimal safe gaps $\Delta_{t,1}$ and $\Delta_{t,2}$ to avoid the collisions at conflict areas are set as $1.5 s$ and $2 s$, respectively. In addition, the vehicle dynamics parameters $a_{\max}$, $a_{\min}$, $v_{\max}$, $v_{\min}$ and $v_c$ are set as $3 m/s^2$, $-5 m/s^2$, $15 m/s$, $0 m/s$ and $10 m/s$ respectively. As suggested in \cite{16}, the maximal searching time of the MCTS method utilized in the distributed strategy is set as $0.1s$, which guarantees the on-line computation in the practical applications.
\par All simulations are carried out on MATLAB R2018a and Visual Studio platform in a personal computer with an i7 CPU and a 16 GB RAM.
\subsection{Performance Analysis under Simple Road Network}
\par
\begin{figure}[h]
\centering
\includegraphics[width=3.2in]{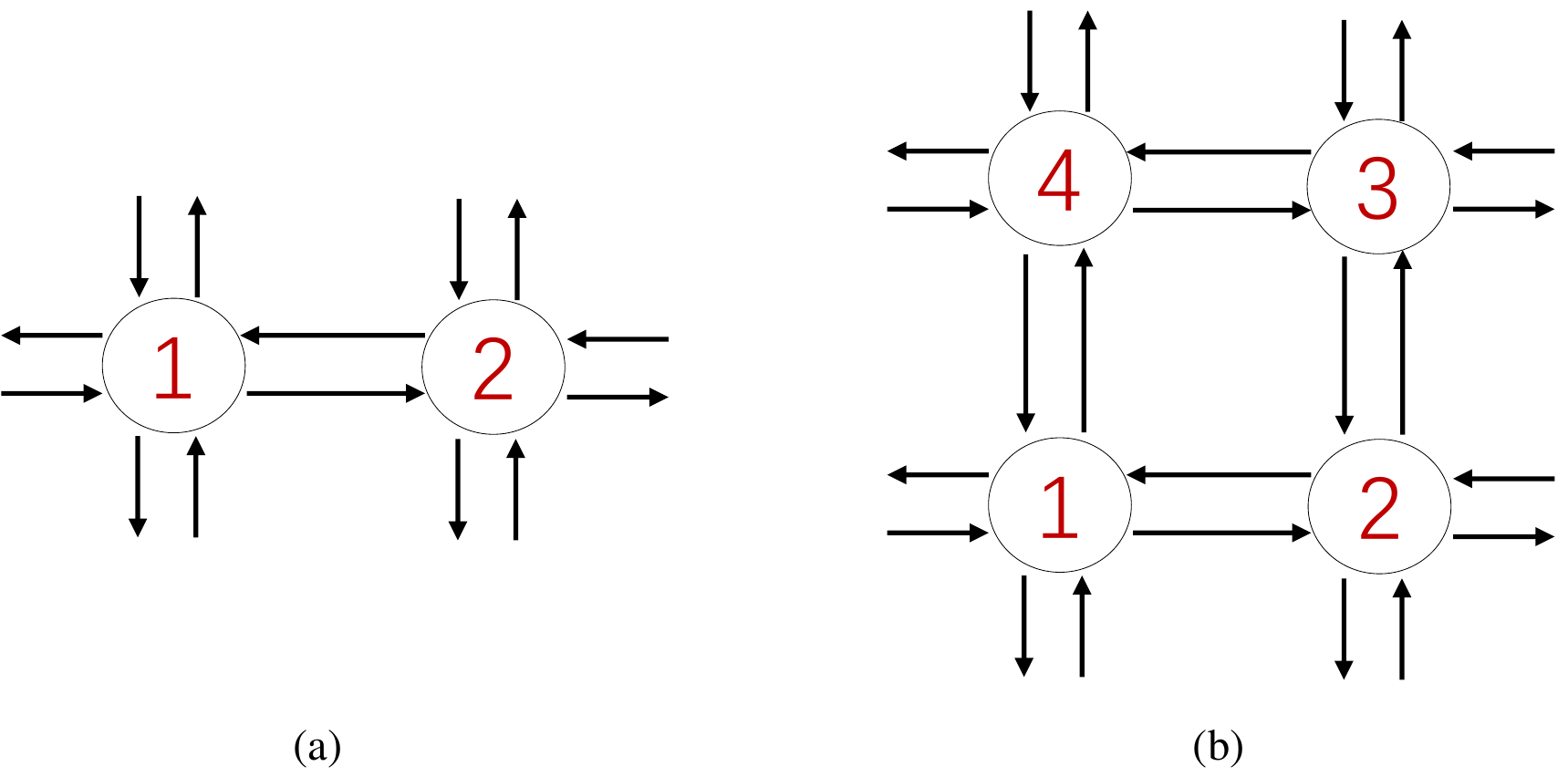}
\caption{The simple but representative road networks used in \emph{Section \Rmnum{4}-B}.}
\label{fig-R4}
\end{figure}

 Considering that it is difficult to obtain the globally optimal solution in a large-scale road network, in this subsection, we first analyze the optimality of the proposed strategy under a simple road network shown in Fig. \ref{fig-R4}(a), and then preliminarily evaluate its coordination performance in terms of improving traffic efficiency under a representative road network shown in Fig. \ref{fig-R4}(b). The distance between adjacent conflict areas is designed as $400m$, and $l_r$ and $l_c$ are set as $100m$ and $200m$ respectively. The time interval between two successive planning procedures $\Delta_T$ is set as $2s$.
\subsubsection{Optimality Analysis} we compare the solution obtained by the proposed distributed strategy with the globally optimal solution obtained by directly solving the MILP problem (\ref{8}) using CVX software with Mosek solver, where the modified branch and bound algorithm is used to handle the binary variables. We call the strategy that can get the globally optimal solution as the optimal strategy. In addition, similar to \cite{14,15,16}, the typical first-in-first-out (FIFO) strategy is also utilized for comparison. Two performance metrics are used in this simulation, i.e., average delay of vehicles passing through the whole road network calculated by (\ref{3}) and average computation time used to get the vehicle sequence. We simulate 10 minutes continuous traffic process for each arrival rate setting. The simulation results are presented in TABLE \ref{T1}.
\begin{table}[h]
\renewcommand{\arraystretch}{1.2}
\centering
\caption{Simulation Results of Optimality Analysis.}
\centering
\begin{tabular}{cccc}
\Xhline{1.2pt}
\bfseries{Arrival Rate} & \bfseries{Cooperative Driving} & \bfseries{Total} & \bfseries{Computation}\\
($veh/h$) & \bfseries{Strategies} & \bfseries{Delay$^{\rm a}$ ($s$)} & \bfseries{Time ($s$)}\\
\Xhline{1.2pt}
\multirow{3}{*}{1200} & FIFO Strategy        & 1.9722  & 0.0024  \\
                      & Optimal Strategy     & 1.3406  & 0.4518  \\
                      & Distributed Strategy & 1.7331  & 0.1     \\ \hline
\multirow{3}{*}{2400} & FIFO Strategy        & 24.7314 & 0.0022  \\
                      & Optimal Strategy     & 2.8081  & 1.0066  \\
                      & Distributed Strategy & 3.1039  & 0.1     \\ \hline
\multirow{3}{*}{3600} & FIFO Strategy        & 65.4656 & 0.0045  \\
                      & Optimal Strategy     & 5.4663  & 75.3943 \\
                      & Distributed Strategy & 5.7381  & 0.1     \\
\Xhline{1.2pt}
\end{tabular}
\label{T1}
\begin{tablenotes}
\item[1] $^{\rm a}$Average delay of vehicles passing through the whole road network.
\end{tablenotes}
\end{table}
\begin{table*}[t]
\renewcommand{\arraystretch}{1.2}
\centering
\caption{Comparison Results of Different Strategies under the Road Network Consisting of Four Intersections.}
\centering
\begin{tabular}{cccccccc}
\Xhline{1pt}
{\bfseries Arrival Rate} & \bfseries Cooperative Driving & \multirow{2}*{\bfseries Total Delay $(s)$}  & {\bfseries Average Speed} & \multirow{2}*{\bfseries Delay 1$^{\rm a}$ $(s)$} & \multirow{2}*{\bfseries Delay 2$^{\rm b}$ $(s)$} & \multirow{2}*{\bfseries Delay 3$^{\rm c}$ $(s)$} & \multirow{2}*{\bfseries Delay 4$^{\rm d}$ $(s)$}\\
$(veh/h)$ & \bfseries Strategies & ~ & ($m/s$) & ~ & ~ & ~ & ~\\
\Xhline{1pt}
\multirow{2}{*}{1200}                                          & FIFO Strategy       & 1.8636                                                    &12.8034                                                       & 0.9111                                                & 0.7454                                                & 0.5946                                                & 1.3006                                                \\
                                                               & Distributed Strategy         & 0.9119                                                    & 13.1093                                                      & 0.5785                                                & 0.4859                                                & 0.3906                                                & 0.4095                                                \\
\hline
\multirow{2}{*}{2400}                                          & FIFO Strategy      & 2.2289                                                    & 12.6885                                                       & 1.0578                                                & 1.0695                                                & 1.0789                                                & 1.2487                                                \\
                                                               & Distributed Strategy         & 1.3356                                                    & 12.9729                                                       & 0.6375                                                & 0.6362                                                & 0.6825                                                & 0.7151                                                \\
\hline
\multirow{2}{*}{4800}                                          & FIFO Strategy       & 24.7228                                                   &8.5098                                                       & 11.9398                                               & 10.5659                                               & 12.2515                                               & 11.3652                                               \\
                                                               & Distributed Strategy         & 3.0898                                                    & 12.4495                                                       & 2.0031                                                & 1.5472                                                & 1.2535                                                & 1.3229
                                                               \\
\Xhline{1pt}
\end{tabular}
\label{Table3}
\begin{tablenotes}
\item[1] $^{\rm a}$Average delay of vehicles at intersection 1. $^{\rm b}$Average delay of vehicles at intersection 2.
\item[2] $^{\rm c}$Average delay of vehicles at intersection 3. $^{\rm d}$Average delay of vehicles at intersection 4.
\end{tablenotes}
\end{table*}
\par The simulation results indicate that the average delay of the proposed distributed strategy is nearly equal to the globally optimal value obtained by the optimal strategy and significantly decreased compared to the typical FIFO strategy. In addition, the average computation time of the distributed strategy is close to that of FIFO strategy, while the computation time of the optimal strategy grows dramatically with the increasing number of vehicles incorporated in the planning procedure. Consequently, we can conclude that the proposed distributed strategy can keep a good trade-off between traffic efficiency and computational complexity in a road network scenario.
\subsubsection{Performance Evaluation} we implement the comparison simulations under a grid of intersections to preliminarily evaluate the coordination performance of the proposed strategy. Note that it is extremely difficult to obtain the globally optimal solution in a large-scale road network scenario, so the optimal strategy is not adopted in this comparison simulation. Three kinds of performance metrics are used in this simulation, i.e., average delay of vehicles passing through the whole road network, average delay of vehicles at each intersection and average speed of passing through the whole road network. We simulate 10 minutes continuous traffic process for each arrival rate setting. The simulation results are shown in TABLE \ref{Table3}.
\par The simulation results indicate that the proposed distributed strategy outperforms the FIFO strategy in terms of all performance metrics, and the advantage increases with the increase of vehicle arrival rate. It demonstrates the promising performance of the proposed strategy in terms of improving traffic efficiency. It is worth emphasizing that, in addition to the advantages in terms of the average delay of passing through the whole road network, the proposed strategy also works better than the FIFO strategy in terms of the average delay at every single intersection and the average speed, which validates that the proposed strategy can effectively alleviate traffic congestions at every single intersection in the road network and achieve traffic balance.
\subsection{Scalability Verification under Complex Road network}
\par In \emph{Section \Rmnum{4}-B}, we have preliminarily evaluated the promising performance of the proposed strategy under a simple but common road network with two or four intersections. In this subsection, we will further discuss the scalability of the distributed strategy in a large-scale and randomized road network that consists of many multi-lane intersections at different places and with random distances from each other.
\par
\begin{figure}[h]
\centering
\includegraphics[width=3.2in]{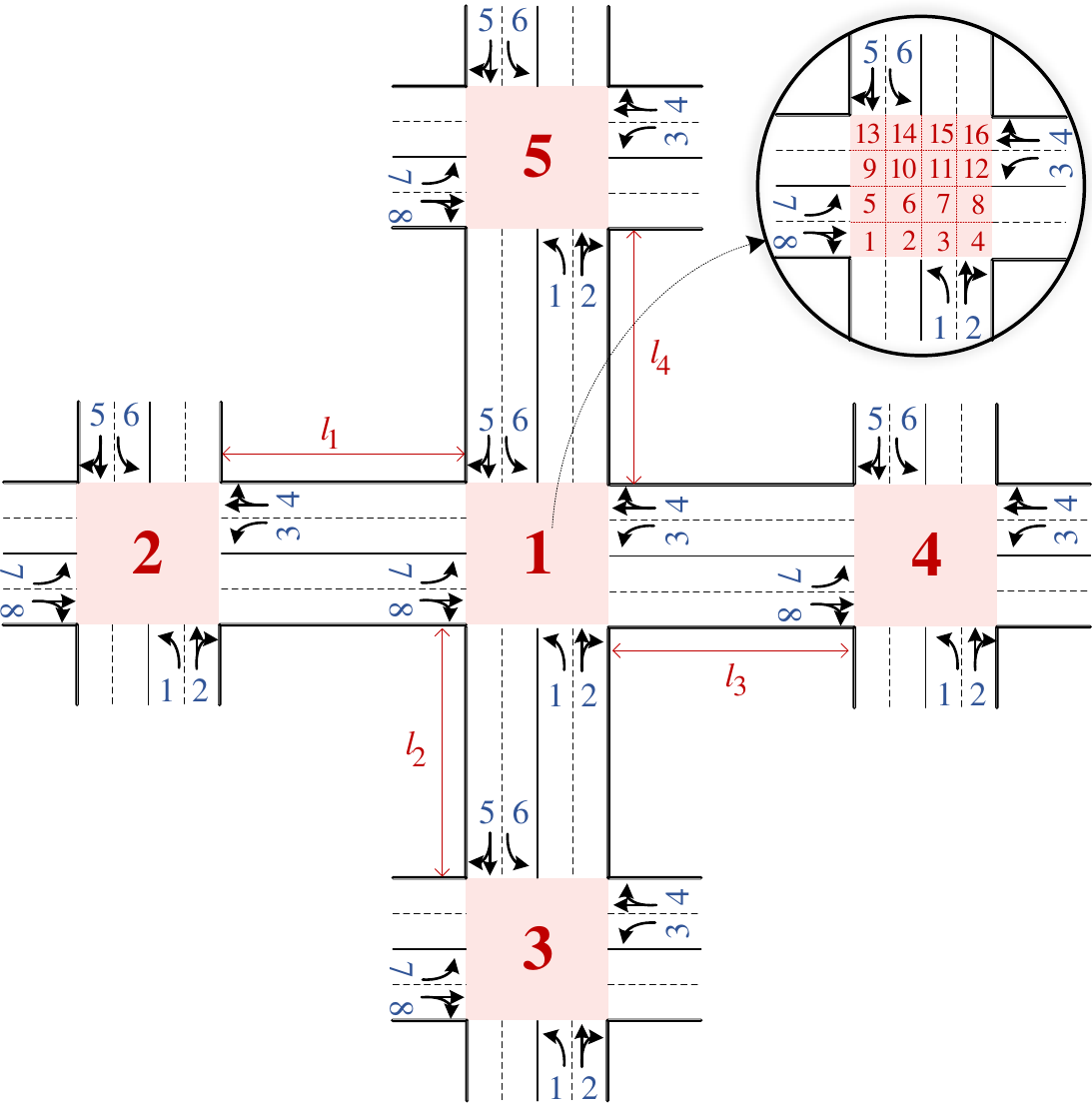}
\caption{A basic and representative node of the large-scale road networks.}
\label{R1}
\end{figure}

 As mentioned, the vehicles far away from the current conflict area have a trivial influence on planning trajectory for the vehicles close to the current conflict area. Therefore, in the proposed distributed strategy, each intersection needs to cooperate with its adjacent intersections and ignore other unconnected intersections in the road network to keep a good balance between computational complexity and traffic efficiency. In this simulation, the scalability of the proposed strategy is verified via evaluating the coordination performance at a representative intersection which has four adjacent intersections, as shown in Fig. \ref{R1}, where $l_i$, $i\in\{1,2,3,4\}$, denotes the distance between intersection $1$ and its adjacent intersections respectively. It is a complex road network that consists of five intersections and with random distances from each other. Each intersection is designed as a multi-lane scenario. Importantly, the designed road network is a basic and representative node when implementing distributed cooperative driving in a large-scale road network consisting of many complex intersections, i.e., the performance of the proposed strategy in a large-scale road network can be reflected by that in this designed scenario.
\par In this simulation, we design a randomized network in a rather general scenario where the intersections are with random distance from each other, i.e., $l_1=250m$, $l_2=300m$, $l_3=350m$, $l_4=400m$. Instead of a single-lane scenario, each intersection is with eight input lanes. The arrival rates of vehicles entering the road network are designed as 2400, 4800, 7200, and 9600 $veh/h$ to further evaluate the scalability especially when there is a larger vehicle density in the road network. We set $l_c$ of each intersection as $200m$, and $l_r$ on each leg of the corresponding intersection is set as $(l_i-l_c)$, where $i\in\{1,2,3,4\}$. The time interval between two successive planning procedures $\Delta_T$ is set as $2s$. There are three kinds of performance metrics selected in the simulations, i.e., average delay at intersection 1, average delay of vehicles passing through the whole road network, and average speed of vehicles passing through the whole road network.
\par In the rest of this subsection, we first incorporate various state-of-the-art strategies promoted for isolated intersection into the proposed distributed mechanism to determine which performs better in a multi-intersection road network. Then, various settings are tested to validate the performance of the prediction-based coordination strategy designed in the distributed mechanism to reflect the advantages of assuring cooperation between adjacent areas.
\subsubsection{Comprehensive Evaluations on Different Strategies with the Proposed Distributed Mechanism} the FIFO strategy, dynamic resequencing (DR) strategy \cite{30}, and Monte Carlo tree search (MCTS) strategy \cite{16,24} are the state-of-the-art strategies promoted for planning vehicle sequence at an isolated intersection. In the proposed distributed strategy (DS), we adopted MCTS as the coordination algorithm to solve each sub-problem. In this simulation, FIFO and DR are also incorporated into the proposed distributed mechanism as the methods to solve each sub-problem and compared with DS. The results under various traffic demand settings are shown in TABLE \ref{TR1}.
\begin{table}[h]
\renewcommand{\arraystretch}{1.2}
\caption{Comparison Results of Different Strategies in Road Networks.}
\centering
\begin{tabular}{ccccc}
\Xhline{1pt}
\begin{tabular}[c]{@{}c@{}}\bfseries{Arrival Rate}\\ ($veh/h$)\end{tabular} & \bfseries{Strategies} & \begin{tabular}[c]{@{}c@{}}\bfseries{Delay 1$^{\rm a}$}\\ ($s$)\end{tabular} & \begin{tabular}[c]{@{}c@{}}\bfseries{Total delay$^{\rm b}$} \\ ($s$)\end{tabular} & \begin{tabular}[c]{@{}c@{}}\bfseries{Average} \\\bfseries{speed} ($m/s$)\end{tabular} \\
\Xhline{1pt}
\multirow{3}{*}{2400}                                           & FIFO       & 3.7083                                                & 9.3018                                                     & 11.4111                                                        \\
                                                                & DR         & 1.4401                                                & 3.8009                                                     & 12.1099                                                        \\
                                                                & DS         & 1.1382                                                & 3.6625                                                     & 12.1913                                                        \\
\hline
\multirow{3}{*}{4800}                                           & FIFO       & 8.1906                                                & 25.7099                                                    & 9.5021                                                         \\
                                                                & DR         & 3.1599                                                & 8.5948                                                     & 11.5071                                                        \\
                                                                & DS         & 1.7641                                                & 6.6179                                                     & 11.7431                                                        \\
\hline
\multirow{3}{*}{7200}                                           & FIFO       & 22.4209                                               & 48.8589                                                    & 8.3771                                                         \\
                                                                & DR         & 4.0319                                                & 12.2744                                                    & 10.9886                                                        \\
                                                                & DS         & 2.5104                                                & 9.9962                                                     & 11.2685                                                        \\
\hline
\multirow{3}{*}{9600}                                           & FIFO       & 33.5967                                               & 67.8842                                                    & 6.7781                                                         \\
                                                                & DR         & 7.0261                                                & 21.1602                                                    & 9.9586                                                         \\
                                                                & DS         & 5.4461                                                & 17.9007                                                    & 10.3111                                                        \\
\Xhline{1pt}
\end{tabular}
\label{TR1}
\begin{tablenotes}
\item[1]$^{\rm a}$Average delay of vehicles at intersection 1.
\item[2]$^{\rm b}$Average delay of vehicles passing through the whole road network.
\end{tablenotes}
\end{table}
\par Based on the simulation results, we can find that incorporating MCTS into the distributed mechanism (i.e., DS) has better performance than the comparison strategies under various traffic demand settings in terms of average delay at intersection 1, average total delay, and average speed. It worth being noticed that the advantages of DS become more significant with the increase of vehicle density. It indicates that MCTS outperforms the candidate strategies for coordinating the vehicles around every single conflict area when implementing distributed cooperative driving and we can see that DS can also work well in a large-scale and randomized road network even with a large vehicle density.
\subsubsection{Comprehensive Evaluations on the Proposed Prediction-Based Strategy Designed in Distributed Mechanism} in this simulation, the pure decentralized strategy (PDS), which schedules the vehicles around different conflict areas independently while no coordination between adjacent conflict areas, is adopted in the comparison simulations to verify the performance of the prediction-based strategy designed in DS. To guarantee fairness, the MCTS is also introduced into PDS with the same settings to independently schedule the vehicles within the control range of each individual intersection. The simulation results under various traffic demand settings are shown in TABLE \ref{TR2}.
\begin{table}[h]
\renewcommand{\arraystretch}{1.2}
\caption{Simulation Results for Evaluating Prediction-based Strategy.}
\centering
\begin{tabular}{ccccc}
\Xhline{1pt}
\begin{tabular}[c]{@{}c@{}}\bfseries{Arrival Rate}\\ ($veh/h$)\end{tabular} & \bfseries{Strategies} & \begin{tabular}[c]{@{}c@{}}\bfseries{Delay 1}\\ ($s$)\end{tabular} & \begin{tabular}[c]{@{}c@{}}\bfseries{Total delay} \\ ($s$)\end{tabular} & \begin{tabular}[c]{@{}c@{}}\bfseries{Average} \\\bfseries{speed} ($m/s$)\end{tabular} \\
\Xhline{1pt}
\multirow{2}{*}{2400} & PDS & 1.7167 & 4.8766  & 11.9905 \\
                      & DS  & 1.1382 & 3.6625  & 12.1913 \\
\hline
\multirow{2}{*}{4800} & PDS & 2.4106 & 7.8566  & 11.5483 \\
                      & DS  & 1.7641 & 6.6179  & 11.7431 \\
\hline
\multirow{2}{*}{7200} & PDS & 3.3085 & 11.9501 & 11.0063 \\
                      & DS  & 2.5104 & 9.9962  & 11.2685 \\
\hline
\multirow{2}{*}{9600} & PDS & 6.0441 & 18.7891 & 10.1845 \\
                      & DS  & 5.4461 & 17.9007 & 10.3111 \\
\Xhline{1pt}
\end{tabular}
\label{TR2}
\end{table}
\par According to the simulation results, the DS which utilizes the information of vehicles outside the control range to implement predictive coordination has distinctively better performance than PDS in terms of average delay at intersection 1, average total delay, and also average speed. It demonstrates the advantages of assuring cooperation between adjacent areas in network-wide cooperative driving and further verifies that DS can keep a good balance between computational complexity and traffic efficiency in complex road networks.
\subsection{Sensitivity Analysis}
\par In \emph{Section \Rmnum{4}-C}, it preliminarily indicates that DS has the potential to generalize to various road networks. In this subsection, we focus on sensitivity analysis to discuss how the coordination performance varies with respect to some key factors including road network geometry and parameter settings of DS.
\par The road network geometry directly determines how to divide the road network when implementing distributed cooperative driving. In this regard, the sensitivity analysis with respect to road network geometry can be performed by that with respect to the road network division. As shown in Fig. \ref{fig6}, to implement the proposed distributed strategy, two critical measures need to be determined in the division, i.e., the range of intersection area $l_c$ and the range of road segment area $l_r$ for each conflict area. Generally, $l_c$ can be directly determined by the limited reliable communication range of V2I. Therefore, we present the sensitivity analysis firstly with respect to $l_r$, which determines how much information outside the control range is used for predictive planning. In addition, in the proposed distributed mechanism, we adopt a time-driven rolling horizon optimization mechanism to decompose the large-scale planning problem in the temporal dimension, i.e., the planning procedure is triggered at a regular time interval $\Delta_{T}$. Thus, the time interval between consecutive planning procedures $\Delta_{T}$ is another key parameter in the distributed strategy, so that we also implement sensitivity analysis with respect to this parameter.
\par In this simulation, Fig. \ref{R1} is selected as the simulation scenario again. We set the distance between adjacent conflict areas as $l_1=l_2=l_3=l_4=400m$, and the length of the intersection area is designed as $l_c=200m$. The arrival rate of vehicles is $6000 veh/h$. DS is compared with PDS in terms of the following two representative performance metrics: average delay at intersection 1 and decreased rate $\eta$ of average delay at intersection 1, i.e.,
\begin{equation}
\eta=\frac{J_{PDS}-J_{DS}}{J_{PDS}} \times 100 \%,
\end{equation}
\par\noindent where $J_{PDS}$ and $J_{DS}$ denote the average delay at intersection 1 when using PDS and DS, respectively.
\subsubsection{Sensitivity Analysis with Respect to $l_r$} 
\begin{figure}[h]
\centering
\includegraphics[width=3in]{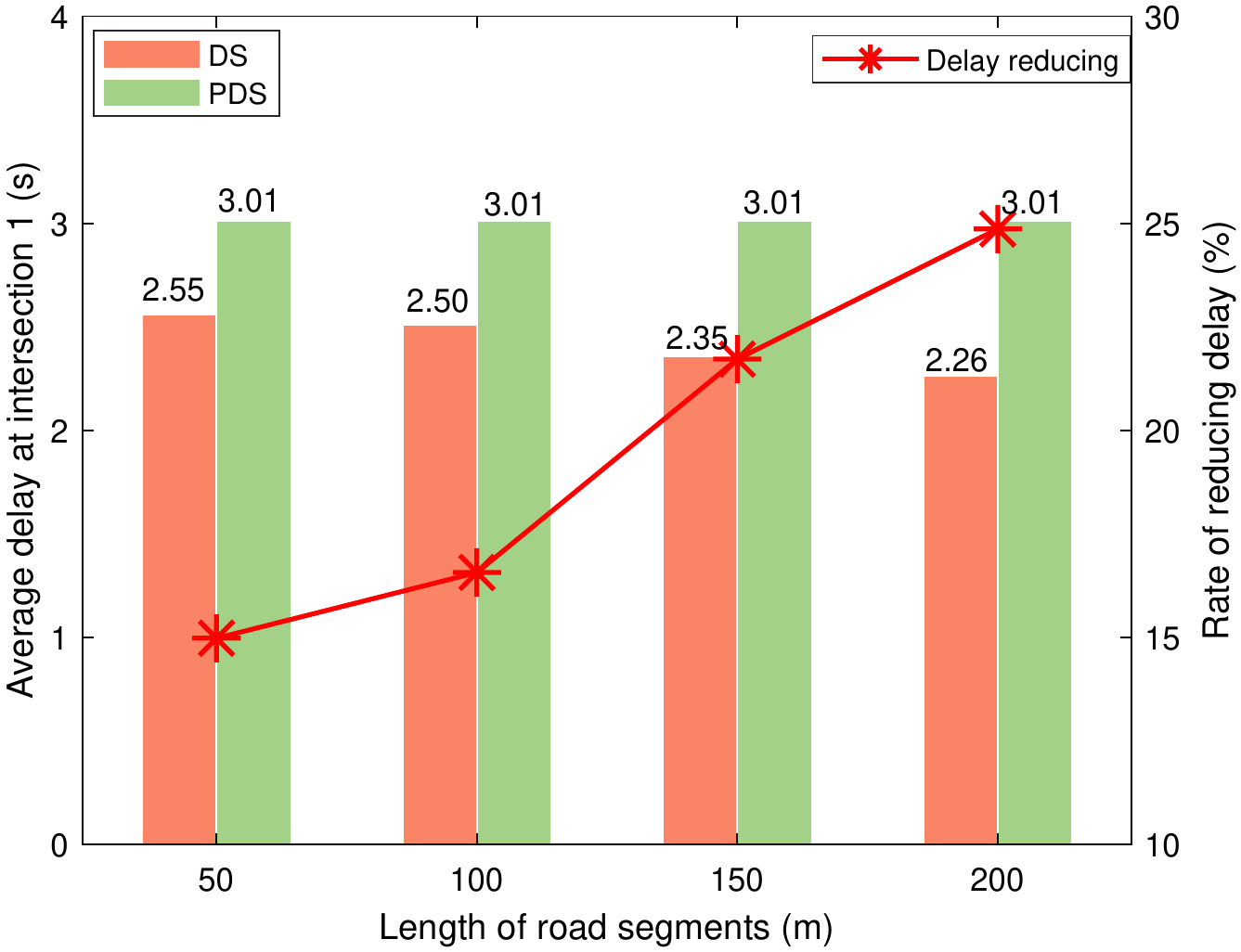}
\caption{The reduction of average delay under different lengths of road segments $l_r$ compared to pure decentralized strategy.}
\label{Fig1-R}
\end{figure}

in this simulation, the time interval between two consecutive planning procedures is set as $\Delta_T=4s$. The length of the road segments $l_r$ is varied as $50m$, $100m$, $150m$ and $200m$, and the comparison results are shown in Fig. \ref{Fig1-R}.
\par It can be observed that the longer the road segment considered into each sub-problem is, the more significant the performance improvement of DS produces, as more information of vehicles outside the control range is considered in the corresponding sub-problem to implement predictive coordination. The performance sensitivity with respect to $l_r$ indicates that we should consider longer road segments as much as possible when network geometry and computational resources permit.
\subsubsection{Sensitivity Analysis with Respect to $\Delta_T$} 
\begin{figure}[h]
\centering
\includegraphics[width=3in]{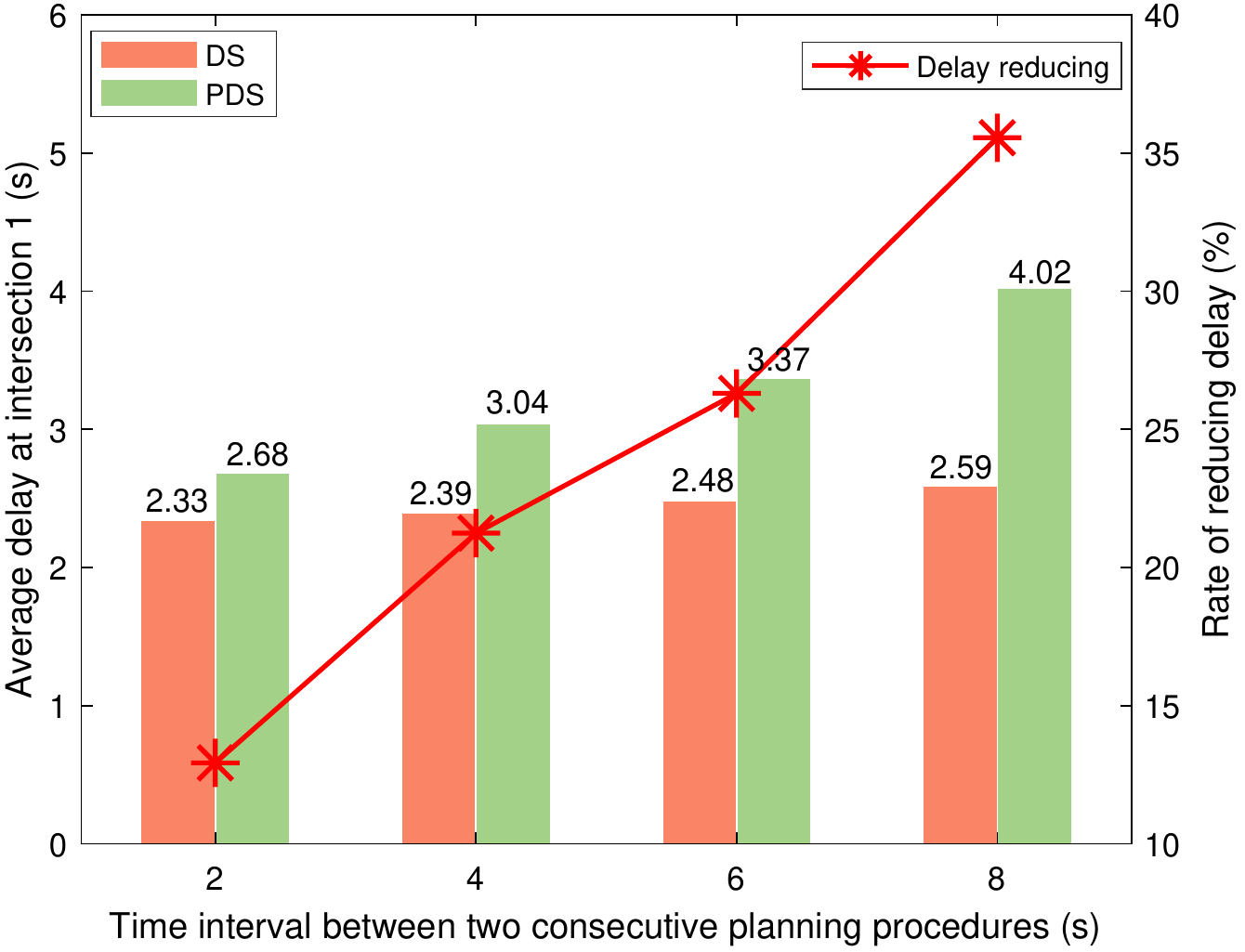}
\caption{The reduction of average delay under different time intervals between consecutive planning procedures compared to pure decentralized strategy.}
\label{Fig2-R}
\end{figure}

in this simulation, the length of the road segment considered into each sub-problem is set as $l_r=200m$. The time interval between two consecutive planning procedures $\Delta_T$ is varied as $2s$, $4s$, $6s$ and $8s$, and the comparison results are shown in Fig. \ref{Fig2-R}.
\par According to the simulation results, it can be found that for the proposed DS, the average delay increases along with the time interval between two consecutive planning procedures $\Delta_T$, as the trajectory of vehicles can be updated more timely when $\Delta_T$ is smaller. In addition, the larger the time interval $\Delta_T$ is, the more significant the performance improvement of DS produces, as each sub-problem of DS can earlier utilize the information of vehicles outside the control range. It should be noted that, in practical applications, $\Delta_T$ is usually not very small to save computational resources and guarantee the driving stability of vehicles. Therefore, according to the performance sensitivity with respect to $\Delta_T$, we can find an appropriate value of this parameter under jointly considering the coordination performance, computational resources, and driving stability of vehicles.
\section{Conclusions}
\par In this work, a distributed strategy is proposed to attack the challenging problem of network-wide cooperative driving. In the proposed strategy, the large-scale planning problem is sequentially decomposed to reduce computational complexity, and the prediction-based coordination between adjacent areas is implemented to guarantee traffic efficiency. Multiple simulations jointly demonstrate that the proposed strategy can approach a sufficiently optimal solution within limited computation time and is able to generalize to different road networks under various traffic demand settings.
\par In future work, some interesting aspects are also worth being further studied. For instance, pedestrians can be considered in intersection management in urban road networks \cite{8883362}; the scenarios where the human-driven vehicles coexist with CAVs \cite{LAI2020102874} and the scenarios in freeway systems can also be discussed.
\bibliographystyle{IEEEtran}
\bibliography{IEEEabrv,IEEEexample}
\end{document}